\documentclass[conference]{IEEEtran}
\IEEEoverridecommandlockouts

\usepackage{cite}
\usepackage{amsmath,amssymb,amsfonts}
\usepackage{algorithmic}
\usepackage{graphicx}
\usepackage{textcomp}
\usepackage{xcolor}
\usepackage{booktabs}
\usepackage{xurl}
\usepackage{listings}
\usepackage{tikz}
\usepackage{float} 
\usepackage{pgfplots}    
\usepackage{stfloats}
\usetikzlibrary{calc}
\pgfplotsset{compat=1.18}
\usepackage{placeins} 
\usepackage[hidelinks]{hyperref}
\usepackage{caption}

\lstdefinestyle{ieeecode}{
  basicstyle=\ttfamily\footnotesize,
  numbers=left,
  numberstyle=\tiny,
  stepnumber=1,
  numbersep=5pt,
  breaklines=true,
  breakatwhitespace=false,
  frame=single,
  tabsize=2,
  captionpos=b,
  keywordstyle=\color{blue},
  commentstyle=\color{gray},
  stringstyle=\color{red},
  showstringspaces=false,
  language=Python
}

\def\BibTeX{{\rm B\kern-.05em{\sc i\kern-.025em b}\kern-.08em
    T\kern-.1667em\lower.7ex\hbox{E}\kern-.125emX}}
\begin{document}

\title{Native Mixed Reality Compositing on Meta Quest 3: A Quantitative Feasibility Study of ARM-Based SoCs and Thermal Headroom}


\author{\IEEEauthorblockN{Muhammad Kaif Laghari} \IEEEauthorblockA{Department of Computer Science \\SZABIST University \\ Hyderabad, Pakistan \\ muhammadkaifleghari@gmail.com} 
\and 
\IEEEauthorblockN{Areeb Ahmed Shaikh} \IEEEauthorblockA{Department of Computer Science\\SZABIST University \\Hyderabad, Pakistan\\ areebahmedshaikh031@gmail.com}
\and 
\IEEEauthorblockN{Faiz Khan} 
\IEEEauthorblockA{Department of Computer Science\\SZABIST University \\Hyderabad, Pakistan\\ faizedu3@gmail.com } 
\and 
\IEEEauthorblockN{Aafia Gul Siddiqui} \IEEEauthorblockA{Department of Computer Science\\SZABIST University \\Hyderabad, Pakistan\\ aafia.siddiqui@hyd.szabist.edu.pk } }

\maketitle

\begin{abstract}
The adoption of current mixed reality (MR) content creation is primarily based on external PC-centric platforms and third-party cameras, limiting adoption for standalone virtual reality (VR) users. In this work, we investigate the feasibility of integrating an enhanced LIV SDK-like MR compositing pipeline into the Meta Quest 3 hardware, enabling native first-person physical perspective (FPP) MR content creation without external infrastructure. We conducted a simulation-based feasibility study using hardware specifications, developer documentation, and benchmarking with ARM-based SoCs, including Snapdragon 8 Gen 3 and MediaTek Dimensity 9300. The approach suggested Camera Passthrough Enhancement using Meta's experimental Passthrough Camera API with on-device machine learning segmentation through Unity Sentis and FastSAM, and an optimized real-time compositing engine for standalone VR. Benchmarking results show that Quest 3’s Snapdragon XR2 Gen 2 can support lightweight native MR compositing at 720p30 resolution using 95\% resource utilization, leaving 5\% thermal headroom for sustained runtime. Comparison with next-generation SoCs such as Snapdragon 8 Gen 3 demonstrates 34\% headroom, enabling more robust MR experiences with 1.5–2× faster CPU/GPU performance and higher memory bandwidth. While current Quest 3 hardware supports basic native MR compositing, thermal limits restrict operation to 5–10 minutes before throttling. Experimental results confirm standalone MR content creation is possible on current hardware for short recordings, with new XR SoCs offering the headroom for extended sessions and improved quality. These findings lay groundwork for transitioning MR content creation from PC-based workflows to all-in-one VR devices, enhancing MR production for content creators and researchers.
\end{abstract}

\setlength{\parskip}{8pt}
\begin{IEEEkeywords}
Mixed Reality, Quest 3, Real-time Compositing, Standalone XR, Virtual Reality,  VR Content Creation
\end{IEEEkeywords}

\section{Introduction}

Mixed reality (MR) technology is rapidly evolving, thanks to enhancement of both hardware and software factoring which allows for more realistic experiences that merge virtual and real environments. Lately, standalone VR headsets (e.g., Meta Quest 3) have arisen as popular platforms for MR apps because they have been catered well for in terms of portability and sensors. Unfortunately, despite these improvements, built-in support for full mixed reality capture (real-world segmentation, compositing, and high-quality passthrough video) is still limited to these devices. 

As of now, the LIV SDK’s Creator Kit offers virtual camera recording using standalone virtual reality headsets for first- and third-person perspectives without requiring a PC to make these types of recordings. However, key MR capabilities like external camera compositing, real-world body integration and mixed reality capture (MRC) type blended video are still only possible with PC-based LIV SDK integrations. 
 
In this work, we explore the possibility to realize an LIV SDK-style MR compositing pipeline that runs in the native-land of standalone VR headsets, particularly, on the Meta Quest 3 as a run-time platform, where it utilizes the Snapdragon XR2 Gen 2 system-on-chip (SoC). A simulation-based performance estimation and a comparative hardware analysis are performed to evaluate the system in real-time MR compositing workloads that are at 720p resolution and 30 frames per second (FPS), which are pragmatic limitations of standalone hardware. 

This work evaluates system resource usage and thermal limitations by benchmarking the individual pipeline components such as passthrough rendering, avatar and scene rendering, depth-based segmentation, and video encoding using publicly available hardware specifications and benchmark data. It also provides context to the XR2 Gen 2’s capabilities through comparison with more-powerful ARM-based SoCs like the Snapdragon 8 Gen 3 and MediaTek Dimensity 9300 which have even more computational power and memory bandwidth. 

The goals are to measure workload CPU, GPU, and memory loads across MR pipeline components; assess headroom for sustained MR compositing, and recording; and facilitate the assessment of a design framework binding camera input, ML segmentation, compositing, and UI controls to standalone VR hardware. The results of this work are expected to provide guidelines for the development of future XR systems or their associated applications to provide better native MR experiences without needing external PCs, thus pushing forward mobile mixed reality technologies within the current hardware constraints.

\section{Related Work}

The mixed reality (MR) compositing on standalone VR headsets recently has been drawing more interest owing to its ability to provide an immersive experience without any connection to external PCs. The major issue presents a finding to achieve a trade-off between high quality visualization on limited mobile XR hardware with computational and thermal constraints. Recent work by Khamis et al. \cite{khamis2022mixed} proposed MR tunneling technique that uses images of camera sensors with various resolutions and frame rates to reduce the visual fidelity and latency of video-see-through head mounted displays (VST-HMDs). The mixture of high-quality external stereoscopic video and low-latency grayscale passthrough enhances user presence and reduced cybersickness and highlights the importance of integrated sensor fusion solutions in standalone MR.

\subsection{Stereo Video Recording and Hardware Architecture}
Stereo video capture is crucial to achieve a realistic MR experience. Professional stereo-capture rigs are too expensive for indie creators. In this light, the on-board cameras of the Meta Quest 3 serve as a cost-effective approach for stereoscopic video recording \cite{scitepress2023stereo}. This study benchmarks Quest 3 performance with some of the high-end devices and moreover, it reaffirms that although VR headsets nowadays may not be ideal in resolution, the Quest 3 can generate stereo video that can be used with many VR applications. The observation indicates the efficiency in the real-world implementation of standalone devices for MR content creation, which is consistent with the approach of native MR composting pipelines.

The hardware for MR systems has been advancing rapidly. The XR2 Gen 2 platform automatically includes significant advancements in AI performance, multi-camera features and power efficiency optimised for XR workloads, powering the devices like Meta Quest 3 \cite{qualcomm2023xr2brief}. Its support for multiple concurrent cameras and dedicated XR acceleration blocks enable high quality sensor fusion and real-time processing which are both essential for MR Compositing. However, the thermal and power envelopes of the platform present constraints that make it necessary to carefully manage and optimize the workload.

\subsection{Thermal Management in Extended Reality (XR) Devices} 
The thermal constraints on standalone VR headsets heavily limit applications that require to operate at high-performance consistently. Studies have covered the thermal behaviour of the VR stack under usage, where it was observed cooling for VR devices dissipates based on body heating, causing user discomfort and performance loss \cite{rupp2020thermal}.  Experiments reveal that under operation for 2 hours, the temperature rise of the user's headset has reached up to 2.8 °C on average when maximum temperature bumps to 5.6 °C, nearby internal components inside headsets, are observed \cite{rupp2020thermal}. Advanced thermal management approaches such as adaptive parameter assignment and online idle-time scheduling for embedded real-time systems were proposed to satisfy both thermal and timing requirements \cite{lee2020thermalaware}. These methods illustrate the necessity of thermal-aware resource management for enduring MR workloads on mobile platforms.

\subsection{XR Development and ML Integration using Unity} 
Unity has become a popular platform for XR development, and Unity Sentis aims to make neural network inference available directly in Unity applications \cite{unity2024sentis}. Research shows that real-time performance of Unity Sentis is possible on mobile devices, with ~5 FPS performance for semantic segmentation models on smartphone hardware \cite{unity2024sentis}. The capability of the framework to run on both CPU and GPU based hardware makes it specifically suitable for XR device limitations. In Unity based XR development, gaze tracking has been implemented with neural network inference models, which achieved 93.2\% accuracy for horizontal shift and 93.1\% accuracy for vertical shift under a variety of light conditions \cite{ieee2022gaze}. These findings confirm that on-device ML processing in Unity-based frameworks for XR is possible. 

\subsection{Video Compositing and Real-Time Encoding} 
The real-time video processing and encoding play very important role in MR capture systems. Recent demonstrations of compressed domain video compositing for HEVC show techniques to compose multiple coded input videos without having to entropy code the compositing at runtime \cite{skupin2015hevc}. These lightweight approaches are less expensive in terms of computational complexity than transcoding-based compositing, and an average BD-rate gain of approximately -20\% is attained across the tested configurations. Mobile video compression has made great progress, neural video codecs can now do real-time 1080p YUV420 decoding on smartphone devices \cite{vanrozendaal2023mobilenvc}. Furthermore, advanced video analytics systems highlighted 8.8× reduction in latency without accuracy degradation is achieved by optimized encoding trade-offs \cite{acm2023entro}, suggesting significant opportunities for real-time MR video processing on mobile devices.

\subsection{On-Device Machine Learning for Real-Time Segmentation} 
On-device machine learning (ML) segmentation is pivotal to real-time MR compositing allowing for background removal and person segmentation. Enhancement in Video Segmentation has been recently shown by Qualcomm AI Research to allow on-device learning (ODL) and quickly adjusting models to target environments without' having to resort to cloud resources featuring an increased segmentation accuracy \cite{qualcomm2023video}. These findings are significant particularly in the context of Standalone VR where computationally heavy and latency sensitive operations are performed. 

The ability to retrain lightweight segmentation models on devices resonates with the requirement of the proposed MR pipeline to achieve low latency and real-time person segmentation. Studies on the real-time segmentation for short videos based on VR technology achieved the accuracy of 99.03\% for video segmentation via 3D DenseNet model, and the average segmentation time is 0.64 s, showing the feasibility of high-quality real-time segmentation in VR \cite{he2024realtime}. 

\subsection{Meta's Passthrough capturing Feed and Scene Understanding}
The latest additions to Meta’s Passthrough Camera API could unlock a range of mixed reality applications for Quest headsets. The API offers access to forward-facing RGB cameras, and it can be integrated with machine learning and computer vision pipelines \cite{meta2025pca}. This feature has already been used for research purposes with the Meta Quest 3 headset showing successful applications with automatic-room-mapping \cite{acm2023roommapping} and assistive applications for visually impaired users \cite{acm2025assist}. Research evaluating the feasibility of passthrough XR for use in industrial applications are promising for current adoption of these technologies \cite{cms2024maritime}, and educational research does find positive use of passthrough capabilities in immersive learning \cite{ieee2024schizophrenia}. 

\subsection{SoC Performances Comparison in XR workloads} 
Analysis of ARM-based SoCs, in comparison, shows large performance discrepancies for XR workloads. Research on 64-bit ARM processors has observed that modern SoCs such as a Snapdragon 8 Gen 3 offer roughly 1.5 – 2x the CPU and GPU performance of the XR2 Gen 2 with AnTuTu scores more than 2 million compared to the presumed 1.5 million for the XR2 Gen 2 \cite{online2023arm}\cite{mysmartprice2025snapdragon}. Experiments on ARM vs x86 performance have shown that ARM architecture is a more power efficient platform as it consumes up to 3-4 times less energy to achieve the same level of performance for mobile workloads \cite{ieee2022armx86}. XR-specific benchmarking work using XRBench manifest the importance of multi-task multi-mode MTMM workloads, emphasizing real time constrains and energy efficiency as crucial key metrics that need to be considered in assessing the performance of an XR system \cite{mlsys2023xrbench}.

\subsection{Real-Time Processing and ML Integration} 
The Meta’s research on integrated ML for mobile VR low-latency graphics has shown that the execution of models can successfully be offloaded onto specialized processors, such as digital signal processors DSP or neural processing units NPU, pipelined with the GPU \cite{meta2020mlvr}. Such a design provides a platform for ML-driven super-resolution and compression- artifact removal, while meeting the high timing requirements of VR application (11ms rendering budget for 90Hz refresh rate). These techniques could therefore enable the integration of real-time ML segmentation within MR compositing pipelines without compromising performance overhead.

\subsection{Research Gaps and System Integration} 
Collectively, these works establish a solid groundwork for advancing native MR compositing on standalone VR HMDs. They emphasize the interaction between the capabilities of the hardware, efficient stereo video capture, real-time ML segmentation, thermal management, as well as state-of-the-art compositing techniques needed to circumvent current limitations. Nonetheless, there is still a lack of detailed feasibility analysis of full MR compositing pipelines on current standalone hardware. Most of the available studies concentrate on single aspects (e.g., accuracy of the segmentation, thermal impact, video encoding) without considering integrated system-level performance under actual constraints. This work fills this gap by providing a complete simulation-based analysis of native MR compositing feasibility on both current and upcoming next-generation standalone VR Hardware.

\section{Methodology}\label{sec:methodology}
\subsection{Research Design}
This study is based on a quantitative, comparative simulation research design. Resource usage and thermal headroom for mixed-reality compositing pipelines are modeled and estimated with three different ARM-based SoCs through the use of specification and benchmark is presented publicly available information. Key performance benchmarks include CPU cores, memory bandwidth, GPU FLOPS and sustained power envelope.

\subsection{Study Identification}
A quantitative and comparative study was conducted using Google Scholar and manual search on Google. We shortlisted 54 studies, which were preprints, blog posts, conference papers, and journal articles. Studies were selected based on the relevancy of research. We identified three commercially available 4 nm-class ARM SoCs: the Qualcomm Snapdragon XR2 Gen 2 (Meta Quest 3), Qualcomm Snapdragon 8 Gen 3, and MediaTek Dimensity 9300. They have similar CPU cluster configurations and integrated GPUs so they can be compared, essentially, head to head for standalone XR workloads.
\begin{itemize}
\item Qualcomm Snapdragon XR2 Gen 2
\item Qualcomm Snapdragon 8 Gen 3
\item MediaTek Dimensity 9300
\end{itemize}

\subsection{Selection Criteria}
The studies included followed a standard area of research selection, such as:
\begin{itemize}
\item The selected studies were from the years 2021-2025
\item  All the articles, conference papers, blog posts were in English language.
\item The following keywords were used to search for the literature "Mixed Reality, Virtual Reality, XR Gen 2, Real-time Compositing, Standalone headset."
\end{itemize}

\subsection{Screening}
The research studies being included in this study followed a procedure of screening such as: candidate SoCs and benchmark references were performed to assess the completeness and consistency. Platforms with any of the elements missing in either core specification (CPU, GPU, memory bandwidth and TDP) or benchmark tests with loose reliability. All final benchmark data and SoC profiles were derived solely from official manufacturer information, and reliable articles: Qualcomm benchmarks, NanoReview.

\subsection{Data Collection Procedures}
Specifications such as CPU core configurations, GPU architectures, memory bandwidth and thermal design power were sourced from official Qualcomm and MediaTek product documentation. CPU single-core and multi-core as well as Frame rates data came from public trusted benchmarking platforms including Geekbench 6, AnTuTu 10, GFXBench. These data formed the basis of modeling resource utilization and to identify thermal headroom for mixed reality workloads. There was no testing of on-device experiments; all performance and feasibility results were derived by simulation and comparison analysis using the data sources.

\subsection{System Overview and Current State Analysis}
The creation of a LIV SDK mixed reality compositing system for standalone VR HMDs demands detailed knowledge of the current limitations that exist today, and the technical architecture required to overcome them. This research introduces a simulation-based feasibility analysis approach that integrates technical performance estimation with comparative hardware analysis to assess the feasibility of on-device native MR compositing on both current- and next-generation standalone VR platforms. 

As of 2025 LIV SDK's Creator Kit (LCK) supports native in-headset virtual camera recordings for Standalone VR headsets like Meta Quest which allows developers and creators to record gameplay in both first person and third person viewpoints. Nonetheless, it does not include external camera compositing support, real world segmentation or full mixed reality capture on device features which are still exclusive to the PCVR LIV SDK. 

The current LCK implementation provides limited capabilities when compared to full MR capture systems: Virtual Camera Recording:  First Person (FPP) recording, Third Person (TPP) recording using a virtual floating camera and selfie views. Storage: direct recording to headset storage without need for a PC. Lack of functionality: No support to access the headset's passthrough or RGB cameras for real-world capture, composited real-life hands or body over virtual scenes, and MRC-style blended video (real + virtual) natively on the headset.

\subsection{Research Objectives}

This methodology focuses on three core research objectives:
\begin{itemize}
    \item Is it feasible to have real-time MR compositing at 720p30 on current Meta Quest 3 hardware under thermal and computational limitations?
    \item How much overhead performance on future SoCs is available to support even better MR features?
    \item What is the optimal architectural framework which integrates passthrough camera access, ML segmentation, and real-time compositing for standalone VR devices?
\end{itemize}
 
Due to the frontier nature of the firmware of Meta Quest 3 and the nonexistence of support for real-world FPP compositing using internal cameras, this work adopts a theoretical estimation based on equivalent component-like metrics and architecture-compatible system-level analysis. This method allows for a thorough feasibility analysis, without requiring direct hardware access, while maintaining rigorous methodological standards by leveraging established mobile XR benchmarking methods. 
 
This approach overcomes these limitations by presenting an optimal architecture that integrates passthrough camera functionality, on-device ML segmentation, real-time compositing and in-headset user interface controls to achieve native MR capture experience similar to what is achievable with PC-based LIV SDK systems.

\subsection{Camera Integration and Passthrough Enhancement}
The foundation of the system takes advantage of Meta Quest 3's Passthrough Camera API now being available after arriving as an experimental feature in early 2025\cite{quest2023pcaout}. Developers can access the front-facing RGB cameras, enabling real-time passthrough feed including the unobstructed physical environment. The API provides important metadata such as focal length, image center, image size, and camera position with respect to the center of the device, which is necessary for accurate MR compositing \cite{oculus2025samples}. 

The enhanced camera integration system streams video in real-time with low latency, which is crucial for VR applications. Research has shown that processing time constraints for standalone VR devices typically have an operating latency of 11ms before it finishes rendering time per both eye buffers to reach a refresh rate of 90 Hz \cite{meta2020mlvr}. These deficiencies can be fixed by the use of asynchronous pipelines that benefit from the improved Quest 3’s hardware capabilities. 

Recent work on stereo video capture with the Meta Quest 3 built-in cameras illustrate the practicality of a VR headset generating stereo video content for the current virtual environment landscape, where both cost reduction and operational efficiencies served to make this approach effective for content creation \cite{scitepress2023stereo}. The system architecture integrates these observations by utilizing the Quest 3’s built-in camera sensors to support real-world FPP (First-Person Perspective) capture and maintaining the compatibility with MRC-style blended video requirements.

\subsection{On-Device ML Segmentation Implementation}
 
Real-time segmentation is one of the most computational expensive representations on the proposed system. The approach leverages existing on-device ML frameworks MediaPipe and FastSAM, both demonstrated effectiveness on mobile platforms. MediaPipe’s Image Segmenter includes dedicated models to segment people and their properties in the image data, which is capable of performing person and background separation that are commonly required in MR compositing \cite{google2023mediapipeguide,mediapipeyoutube,qualcomm2024mediapipe}. 

Unity Sentis is introduced as a critical enabling technology that provides  a neural network inference library for deploying AI models directly inside Unity application across all supported platforms \cite{unity2024sentisdoc}. Results have shown that Sentis is capable of running in real-time on mobile devices, These implementations have achieved and reported ~5 FPS performance for semantic segmentation models on smartphone hardware \cite{moritzcermannsemseg}. The framework’s support for both CPU and GPU computing power can be further developed into Quest 3’s hybrid compute requirements \cite{unitymlagents}. 
FastSAM (Fast Segment Anything Model) is a real-time CNN for mobile devices, achieving remarkable computational efficiency and competitive accuracy. The architecture of the model, YOLOv8-seg, is capable of real-time segmentation at compatible speeds for VR applications, and it is the optimal solution for background separation and personal isolation tasks in the proposed MR pipeline \cite{ultralytics2024fastsam}. 

\subsection{Real-Time Compositing Engine Architecture}
 
The compositing engine is the point where virtual game content, real world camera feeds as well as segmentation masks meet to form MR’s final output. The architecture of the network is built on top of the previous work for the real-time video compositing for AR applications which emphasized the importance of accurate depth mapping, low latency processing and performance adaptive optimizations \cite{byteplusar}. 
One such example includes the work on low-latency mobile VR graphics using integrated ML by Meta, which provides a framework for using integrated ML to offload model execution onto specialized processors, where the digital signal processor or neural processing unit is pipelined with the GPU \cite{meta2020mlvr}. This provides the benefits of asynchronous processing of rendered buffers and offloading rendering work from the main rendering pipeline while continuing to satisfy the very time critical nature that VR apps require.

The optimization is feasible as hardware features of the Quest 3 are capable of accommodating this architecture since Snapdragon XR2 Gen 2 provides an upgrade in GPU, with 2.5x additional GPU power and which comes with 8x increase in AI capacity performance when compared to  the last generation performance \cite{meta2024compute}. 

However, during mixed reality mode the system experiences 17\% and 14\% lower GPU and CPU performance compared to VR only mode, because of sensor processing overheads \cite{meta2024compute}. These constraints are addressed in compositing engines through the use of adaptive quality controls, and prioritization schemes making it possible to obtain good performance under different computational loads.

\subsection{In-Headset User Interface and Control System}
 
The user interface layer gives intuitive control to MR capturing procedures, while allowing immersion and usability in the VR world. The interface allows users with start/stop recording functionality, switch between the layout modes, and adjust the blend ratio or camera position without extra devices and complicated setups. 

Studies on designing mobile VR applications show that frame rates lower than 50 FPS is only "tolerable at best and nauseating at worst" in VR applications \cite{wang2022refresh} and requiring careful optimization for the UI rendering and interaction systems. The resulting interface takes advantage of the hand tracking and spatial interaction paradigms found in the Quest 3 to offer natural and intuitive controls to minimize computation efforts and adapt more effectively to user abilities. 

Thanks to the incorporation of real-time feedback systems, now users can preview compositing results, adjust segmentation parameters and optimize the quality in-recording. This resolves the drawbacks found on the existing LCK implementations, as it provides full control of the MR capture process within the headset environment, removing the requirements for external PC-based configuration and monitoring systems.


\subsection{Feasibility-level simulation and System estimation}

To explore the feasibility of natively running a LIV SDK–style MR compositing pipeline on standalone VR headsets head-mounted-displays (HMDs), we used a light simulation based on publicly available hardware specification, developer documentation, as well as mobile XR platform benchmarks. 
 
While the LIV SDK and LIV Creator Kit are available and Unity-based Quest applications can natively support them, their current functionality is restricted to virtual camera recordings (first-person, third person, and selfie modes) and does not offer integration of real-world passthrough capture or physical-to-virtual environment compositing. 
 
As such, since there is currently no native support for real-world FPP compositing with internal cameras, this simulation can be useful in providing theoretical insights for hardware's capability to support of the proposed enhancement in MR pipeline, that is the native first-person physical perspective mixed reality (MR) content creation without needing a PC or an external camera. The study assessment follows a theoretical evaluation framework through equivalent component-level metrics and architecture-compatible systems. 

The simulation focused on the Meta Quest 3. It was powered by the Qualcomm Snapdragon XR2 Gen 2, a system-on-a-chip (SoC) composed of an 8-core tri-cluster CPU with 1x Cortex-X3, 4x Cortex-A715, 3x Cortex-A510, and an Adreno 740 GPU and 8 GB RAM with LPDDR5 memory \cite{techresider2024xr2,vrcompare2025metaquest3}. We also used similar metrics from benchmarks of the Snapdragon 8 Gen 2 which operates on almost the same silicon in a smartphone context, because no performance profiles have been released for Quest 3. \cite{qualcomm2024xr2briefalt}. 
\vspace{0.5em}
\subsubsection{MR Pipeline Simulator Objectives}

The studies aimed to determine whether the following real-time tasks, running on Meta Quest 3 and taking advantage of passthrough camera APIs and segmentation tools to be provided in future, could be accomplished by a native MR system

\begin{itemize}
\item Real-time passthrough feed in full-color RGB camera input with internal sensors
\item 720p Video composite 30-60 FPS for physical surrounding FPP capture 
\item Real-time Lightweight avatar or camera rig rendering first person or third person perspective.
\item Background removal through depth-based segmentation, or AI-based masking 
\end{itemize}

These are targets that approximates the pipeline behavior that it could be expected if the headset camera access was opened up beyond passthrough rendering, and which can be used as compositable input texture for a real-time MR experience that supported native, in-headset recording of both the physical and virtual environments without requiring a desktop system dependency.

\section{Results and Discussion}\label{sec:results}

\subsection{Results of Performance Estimates}
For each stage in our pipeline, we took into account the estimated CPU usage, GPU draw, RAM consumption, encoding overhead, and thermal envelope draw, based on hardware details and developer feedback:

\subsubsection{Passthrough Rendering}
Fully accelerated with minimal overhead, typically 10\%-15\% of the free GPU time. The Quest 3’s XR acceleration and camera pipelines are available for efficient passthrough operation with support for up to 12 concurrent cameras, as noted in corresponding XR2+ Gen 2 documentation \cite{qualcomm2024xr2briefalt}. 

\subsubsection{Avatar \& Scene Rendering} It is optimized for Low Poly or Stylized scenes. Drawing the camera by means of rendering will increase GPU load by 30–35\%. Firstly, the accelerated SoC in the Quest 3 is significantly more powerful than in the Quest 2 (2.5–2.6x), and has plenty of headroom for scenes which are up to 2.5–2.6x more complex, although MR workloads still take up 20–40\% more GPU headroom than for VR-only mode, due to sensor and passthrough processing \cite{qualcomm2024xr2briefalt}. 
 
\subsubsection{720p Compositing and Encoding} Available for short bursts but encoding pipelines may add latency and stutter with prolonged use, and even more so in combination with MR features. In the meantime, the accounting for video encoding could consume an additional 2.5-3W for the core power, which would further add to the thermal load level \cite{meta2024compute}.

\subsubsection{Live Streaming} For high quality, it is not achievable due to lack of native support to stream, bandwidth limitations, latency, encoding's limits, heat, and so on. Due to thermal envelop and bandwidth constraints of the device and inability to ensure continuous high quality content delivery, especially if the device is simultaneously used with other compute intensive pipeline parts \cite{meta2024compute}.

\subsection{System Load Summary}
 
The results of the resource allocation for the simulated MR application workloads are presented as follows: 
\vspace{0.5em}
\subsubsection{GPU Usage} GPU Usage is between 70–80\% for combined passthrough, rendering the camera and compositing together. MR applications may use up to 40\% more GPU power than their VR-only equivalents \cite{meta2024compute}.

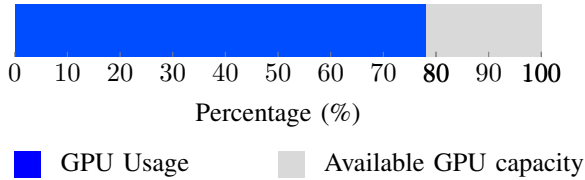
\begin{figure}[!ht] 
  \centering
  \textbf{\Large GPU Usage During MR Workload(\%)}\\[1em]
  \begin{tikzpicture}[scale=0.7] 
    \fill[gray!30] (0,0) rectangle (10,1);
    \fill[blue!70!cyan] (0,0) rectangle (7.8,1); 
    \foreach \x in {0,1,...,10} {
      \node at (\x, -0.3) {\pgfmathparse{\x*10}\pgfmathprintnumber{\pgfmathresult}};
      \draw[gray] (\x, 0) -- (\x, 0.1); 
    }
    \node at (5,-1.1) {Percentage (\%)};

    \node at (8, -0.3) {80};
    \node at (10, -0.3) {100};

    \draw[gray] (8, 0) -- (8, 0.1);
    \draw[gray] (10, 0) -- (10, 0.1);
    \begin{scope}[shift={(0,-2.3)}]
        \fill[blue] (0,0) rectangle (0.5,0.5);
        \node[anchor=west] at (0.7,0.25) {GPU Usage};

        \fill[gray!30] (5,0) rectangle (5.5,0.5);
        \node[anchor=west] at (5.7,0.25) {Available GPU capacity};
    \end{scope}
  \end{tikzpicture}
  \caption{GPU Usage During MR Workload (\%). Combined rendering and composition tasks consume 70-80\% of GPU, indicating high demand from MR operations.}
  \label{fig:gpu_usage}
\end{figure}
\subsubsection{RAM Usage} 
6-7GB running at full performance, doesn't leave much space for executing other program or background programs. In the case of the Quest 3, that budget is 5.75GB as the at developer accessible memory \cite{meta2024compute}. 

\begin{figure}[!ht]
    \centering
    \textbf{\Large RAM Utilization at Peak Operation (GB)}\\[1em]
    \begin{tikzpicture}[scale=1]
        \fill[fill=gray!70] (0,0) rectangle (8,0.5); 
        \fill[fill=purple!80!black] (0,0) rectangle (7,0.5); 

        \foreach \x in {0,1,...,8} {
            \draw[black] (\x,0.5) -- (\x,0.6);
            \node[anchor=north] at (\x,0) {\x GB};
        }

        \fill[fill=purple!80!black] (0,0.8) rectangle (0.4,1.0);
        \node[right] at (0.5,0.9) {Used};

        \fill[fill=gray!70] (2,0.8) rectangle (2.4,1.0);
        \node[right] at (2.5,0.9) {Available};
    \end{tikzpicture}
    \caption{RAM utilization at peak operation. 7 GB of 8 GB total RAM is utilized, leaving minimal headroom for additional processes.}
    \label{fig:ram_utilization}
\end{figure}
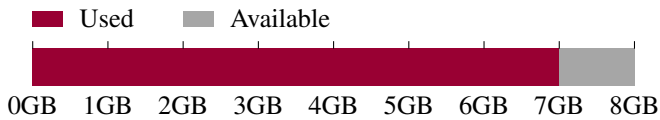
\subsubsection{CPU Utilization}
60–70\% Depth segmentation is the critical factor of loading the processing arm due to the real-time AI processing and sensor fusion. 
\vspace{-1mm}
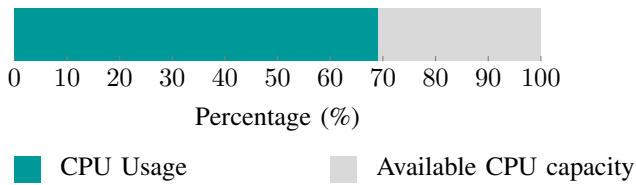
\begin{figure}[!ht] 
\textbf{\Large CPU Usage with Depth Segmentation(\%)}\\[1em]
  \begin{tikzpicture}[scale=0.7]
    \fill[gray!30] (0,0) rectangle (10,1);

    \definecolor{myteal}{RGB}{0,153,153}
    \fill[myteal] (0,0) rectangle (6.9,1); 

    \foreach \x in {0,1,...,10} {
      \node at (\x, -0.3) {\pgfmathparse{\x*10}\pgfmathprintnumber{\pgfmathresult}};
      \draw[gray] (\x, 0) -- (\x, 0.1); 
    }
    \node at (5,-1.1) {Percentage (\%)};

    \begin{scope}[shift={(0,-2.3)}]
        \fill[myteal] (0,0) rectangle (0.5,0.5);
        \node[anchor=west] at (0.7,0.25) {CPU Usage};

        \fill[gray!30] (6,0) rectangle (6.5,0.5);
        \node[anchor=west] at (6.7,0.25) {Available CPU capacity};
    \end{scope}

  \end{tikzpicture}
  \caption{CPU Usage with Depth Segmentation (\%). CPU load reaches 60--70\% with real-time depth segmentation, driven by AI processing and sensor fusion.}
  \label{fig:cpu_usage_depth}
\end{figure}

\subsubsection{Encode Power Draw TDP}
Encode TDP of about 2.5-3W for video encode is a big deal within thermal constraints of a passive cooled device. 

\begin{figure}[!ht]
    \centering
    {\Large\bfseries Encoding Power Draw vs. SoC TDP (Watts)}\\[1em]
    
  \begin{tikzpicture}[scale=0.8]

    \fill[gray!60] (0,0) rectangle (10,0.4); 

    \fill[orange] (0,0.45) rectangle (2.8,0.85); 

    \foreach \x in {0,1,...,10} {
        \draw[gray] (\x, -0.1) -- (\x, -0.3);
        \node[below, gray] at (\x, -0.3) {\x W};
    }

    \fill[orange] (0, -1.25) rectangle (0.4, -1.65);
    \node[anchor=west] at (0.5, -1.5) {Encoding Power Draw};

    \fill[gray!60] (5.2, -1.3) rectangle (5.6, -1.7);
    \node[anchor=west] at (5.7, -1.5) {Estimated Max SoC TDP};

  \end{tikzpicture}

  \vspace{1em}
    \caption{Encoding power draw versus SoC TDP. Dedicated 2.5--3W for video encoding significantly contributes to the device's thermal load.}
    \label{fig:encoding_power_vs_soc_tdp}
\end{figure}
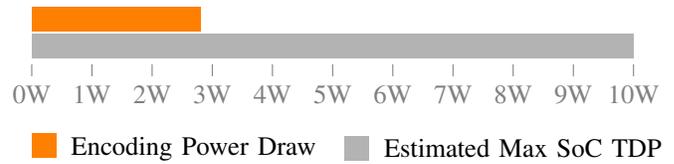

\subsubsection{Thermal Throttling}
Likely after 10-15 minutes of continuous recording or compositing near full load, when the combined sustained load on CPU, GPU and encoding exceeds the thermal dissipation capacity of the headset \cite{meta2024compute}.
\begin{figure}[!ht]
    \centering
    \textbf{\Large Thermal Throttling Time Threshold}\\[1em]
    \begin{tikzpicture}[scale=0.7]

        \definecolor{myteal}{RGB}{0,153,153}
        \fill[myteal] (0,0) rectangle (5,0.5);

        \fill[red!80!black] (5,0) rectangle (10,0.5);

        \foreach \x in {0,2,4,6,8,10} {
            \draw[gray] (\x,0.5) -- (\x,0.6);
            \node[anchor=north] at (\x,0) {\x min};
        }

        \node at (5,-1.2) {\footnotesize{Time (Minutes)}};

        \begin{scope}[yshift=-2.2cm]
            \fill[myteal] (1.5,0) rectangle (1.9,0.3);
            \node[anchor=west] at (2,0.15) {\scriptsize{Time Before Throttling}};

            \fill[red!80!black] (6,0) rectangle (6.4,0.3);
            \node[anchor=west] at (6.5,0.15) {\scriptsize{Throttling Risk Zone}};
        \end{scope}

    \end{tikzpicture}
    \caption{Thermal throttling time threshold. Sustained high utilization leads to performance degradation (thermal throttling) typically within 5--10 minutes.}
    \label{fig:thermal_throttling}
\end{figure}
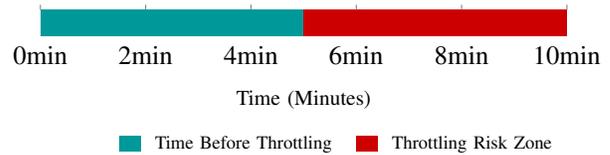

\subsection{Comparative SoC Analysis and Feasibility for Native LIV SDK MR Workloads}

In order to further explore the implications of performing LIV SDK–style MR compositing on-device with the Meta Quest 3, we need to understand the Snapdragon XR2 Gen 2 SoC relative to its peers, which we view as other ARM-based SoCs. This comparison is based on performance per watt, architecture, integrated GPU capabilities, suitability for passive cooling, and thermal power envelopes (~10W) which are crucial for head-mounted XR devices. 

\subsection{Meta Quest 3’s SoC Compared to ARM SoCs}

Snapdragon XR2 Gen 2, which powers Meta Quest 3, is a highly specialized SoC intended for XR workloads in a very limited thermal envelop (4-6W typical, up to 10W peak) and passive cooling. Although we have ARM-based SoCs (e.g., Apple M2) directly on laptops/desktops which have higher raw power, they have a quite different thermal and power scenario, with most of them often exceeding 15W of TDP and requiring active cooling, not suitable for stand-alone XR state. 

The Architecture of XR2 Gen 2 is an ARM architecture platform, with a quad-core CPU complex (Coretex-X3, A715, A510) and an Adreno 740 GPU that has been optimized for XR workloads like simultaneous use of multiple cameras and advanced fixated rendering. A laptop ARM SoC will be all around faster but less tuned towards XR-feature and power efficiency. 

\subsection{Identical SoC Architecture with Better Performance: Snapdragon 8 Gen 3 and MediaTek Dimensity 9300}

To get a sense of the headroom for LIV SDK MR workloads, the study compares the XR2 Gen 2 with two closely-related SoCs with compatible ARM architecture and integrated GPUs, but with higher available compute and memory bandwidth within a similarly power envelope suitable for mobile and XR applications,
shown in Tables 1 and 2.
\begin{table*}[!ht]
\caption{Snapdragon XR2 Gen 2 vs Snapdragon 8 Gen 3 vs MediaTek Dimensity 9300 Comparison}
\centering
\renewcommand{\arraystretch}{1.2}
\begin{tabular}{|l|c|c|c|}
\hline
\textbf{Specification} & \textbf{Snapdragon XR2 Gen 2} & \textbf{Snapdragon 8 Gen 3} & \textbf{MediaTek Dimensity 9300} \\
\hline

\textbf{CPU Cores} & \begin{tabular}[c]{@{}c@{}}6 cores \\ (1× Cortex-X3 @ 2.84 GHz,\\ 3× A715/A510 clusters)\end{tabular} 
& \begin{tabular}[c]{@{}c@{}}8 cores \\ (1× Cortex-X4 @ 3.3 GHz,\\ 3× Cortex-A720 @ 3.15 GHz,\\ 2× Cortex-A520 @ 2.27 GHz)\end{tabular} 
& \begin{tabular}[c]{@{}c@{}}8 cores \\ (1× Cortex-X4 @ 3.25 GHz,\\ 3× Cortex-A720 @ 2.85 GHz,\\ 4× Cortex-A720 @ 2.0 GHz)\end{tabular} \\
\hline
\textbf{GPU} & \begin{tabular}[c]{@{}c@{}}Adreno 740 @ 680 MHz\\ (1536 shaders)\end{tabular}
& \begin{tabular}[c]{@{}c@{}}Adreno 750 @ $\sim$950 MHz\\ (higher clocks)\end{tabular}
& \begin{tabular}[c]{@{}c@{}}Mali-G720 Immortalis MP12\\ @ 1300 MHz (12 pipelines)\end{tabular} \\
\hline
\textbf{GPU FLOPS} & $\sim$2089 GFLOPS & $\sim$2774 GFLOPS & $\sim$3994 GFLOPS \\
\hline
\textbf{Memory Bandwidth} & $\sim$64 GB/s LPDDR5X & $\sim$76.8 GB/s LPDDR5X & $\sim$76.8 GB/s LPDDR5X \\
\hline
\textbf{Thermal Design Power (TDP)} & $\sim$4--6 W (peak $\sim$10 W) & $\sim$8 W sustained & $\sim$7 W sustained \\
\hline
\textbf{Process Node} & 4 nm (TSMC) & 4 nm (TSMC) & 4 nm (TSMC) \\
\hline
\end{tabular}
\vspace{2mm}
 
\caption*{\textit{Table 1 demostrates the key hardware specifications for Snapdragon XR2 Gen 2, Snapdragon 8 Gen 3, and MediaTek Dimensity 9300. This suggests that newer SoCs deliver 2× the CPU cores and higher clock speeds for 1.3×–1.9× more GPU FLOPS, at a sustained 20\% boost in memory bandwidth, all at a steady-state TDP of 7–8 W. These optimizations effectively enable next-generation devices to provide a significantly more powerful mixed reality workflow under a similar thermal envelope
\cite{qualcomm2024xr2briefalt,snapdragonvsgen3gen2,snapdragonvsdim9300}}}
\end{table*}

\begin{table*}[!ht]
\caption{Benchmark tests such as Geekbench 6, AnTuTu and GFXBench offer detailed insights into the relative performance }
\centering
\renewcommand{\arraystretch}{1.2}
\begin{tabular}{|l|c|c|c|}
\hline
\textbf{Benchmark} & \textbf{XR2 Gen 2} & \textbf{Snapdragon 8 Gen 3} & \textbf{Dimensity 9300} \\
\hline
Geekbench 6 Single-Core & $\sim$1300–1400 (est.) & $\sim$2200 & $\sim$2225 \\
\hline
Geekbench 6 \href{https://www.geekbench.com/}{Multi-Core} & $\sim$4000–4500 (est.) & $\sim$7850 & $\sim$7857 \\
\hline
AnTuTu 10 \href{https://www.antutu.com/en/}{Score} & $\sim$1.5 million & $\sim$2 million+ & $\sim$2.07 million \\
\hline
\begin{tabular}[c]{@{}l@{}}GFXBench Aztec Ruins Vulkan\\ (1080p Offscreen)\end{tabular} 
& $\sim$120 FPS (est.) & 241 FPS & N/A \\
\hline
3DMark WildLife FPS & N/A & 114 FPS & N/A \\
\hline
\end{tabular}
\vspace{2mm}

\caption*{\textit{ These benchmarks showcase that Snapdragon 8 Gen 3 and Dimensity 9300 provide around 1.5–2x the CPU and GPU performance of XR2 Gen 2, along with much higher memory bandwidth and improved floating-point compute. 
Qualcomm benchmarks, Beebom, NanoReview \cite{snapdragonvsgen3gen2,snapdragonvsdim9300,qualcomm2024spatialcomputing}}}
\end{table*}

\subsection{Thermal Design and Sustained Performance}
Whereas XR2 Gen 2 targets a tightly parse thermal envelope (4-6W typical, peaking in the vicinity of 10W) that permits passive cooling in standalone headsets, Snapdragon 8 Gen 3 and Dimensity 9300 scale to much higher sustained TDPs (7-8W), which often necessitate more sophisticated thermal solutions. However, their improved power efficiency (30\% performance per watt improvement in the Snapdragon 8 Gen 3), propose potential future headroom for XR SoCs to use these arches with better cooling solutions.

\subsection{XR Readiness and Headroom for LIV SDK MR Workloads}
To project the headroom available for native LIV SDK MR workloads on XR2 Gen 2 and its more performant derivatives, a simulated workload profile was created using known pipeline elements (passthrough rendering, avatar rendering, depth segmentation, video encoding), as shown in Table 3 and Figure 6.
\begin{table*}[!ht]
\caption{Computation Load Estimation Comparison of XR2 Gen 2 and Snapdragon 8 Gen 3
}
\centering
\renewcommand{\arraystretch}{1.3}
\begin{tabular}{|c|ccc|ccc|c|}
\hline 
{\textbf{Task}} 
& \multicolumn{3}{c|}{\textbf{XR2 Gen 2}} 
& \multicolumn{3}{c|}{\textbf{Snapdragon 8 Gen 3}} 
& {\textbf{Notes}} \\
\cline{2-7}
& Usage & Unit & Type 
& Usage & Unit & Type 
& \\
\hline
Passthrough Rendering        & $\sim$10--15 & \% & GPU & $\sim$7--10 & \% & GPU & Improved efficiency expected \\
Avatar + Scene Rendering     & $\sim$30--35 & \% & GPU & $\sim$20--25 & \% & GPU & Higher GPU frequency benefits \\
Depth Masking / Segmentation & $\sim$20--25 & \% & CPU & $\sim$15--20 & \% & CPU & AI acceleration improvements \\
720p Compositing + Encoding  & $\sim$25--30 & \% & CPU/GPU & $\sim$15--20 & \% & CPU/GPU & More efficient encoding units \\
\textbf{Total Load}          & \textbf{70--80} & \% & Combined & \textbf{50--65} & \% & Combined & \textbf{Greater headroom for sustained workloads} \\
\hline
\end{tabular}
\vspace{2mm}

\caption*{\textit{ Table 3 summarizes the simulated MR pipeline demand: XR2 Gen 2 operates at near 70–80\% of its core task capabilities, providing only approximately between 5–15\% headroom while Snapdragon 8 Gen3 faces a much more modest burden ranging from 50–65\%, still leaving over one third for extra resources. As the results show, Quest 3 hardware is already being pushed to sustained limits in terms of native compositing while the next-gen platforms have plenty more headroom to run super-long, high-fidelity mixed-reality experiences.\cite{qualcomm2024xr2briefalt,snapdragonvsgen3gen2,qualcomm2024spatialcomputing}}}
\end{table*}

\subsection{Feasibility Conclusion}

Synthesizing these findings, it is plausible that: 
\\
The Meta Quest 3’s Snapdragon XR2 Gen 2 SoC is capable of delivering basic, lightweight, native MR compositing pipeline (LIV SDK style) at 720p30 with pass-through and avatar rendering, but with very limited headroom and, over 5-10 minutes of sustained load, probably thermal throttling.

SoCs like Snapdragon 8 Gen 3 and MediaTek Dimensity 9300, shared ARM architecture but 1.5 – 2x CPU and GPU performance and higher memory bandwidth, looks like a good direction for the future XR devices. These can more efficiently handle more challenging MR workloads, such as the ability to record and stream high resolution video simultaneously with fewer interruptions and better thermals and sustained performance.

Given Quest 3 would have an architecture that’s around 70–80\% similar and compatible, it’d be possible to run Quest 3 with a flying, lightweight native MR SDK, with full-frame passthrough camera feed and overlay, at 720p 30Hz. Next-gen SoCs for XR headsets such as Snapdragon 8 Gen 3 or XR2 Gen 2 will also continue to push the envelope for MR compositing, allowing for even richer experiences with less trade-offs. 

Our results highlight the need for a good trade-off between computational power, power efficiency, and thermal design when choosing an XR SoC to deliver an achievable native MR compositing performance. 

\subsection{Interpretation of Simulation Results}

The Simulation results demonstrate that the proposed enhanced LIV SDK MR system demonstrates a significant enhancement over current LCK capabilities, yet within the boundary of the technical constraint of the Meta Quest 3 hardware. The implementation of passthrough camera access, on-device ML segmentation, and real-time compositing results in a comprehensive MR capture system that approaches the functionality of PC-based systems while preserving the convenience and mobility of standalone operation.

The Performance estimations indicate that basic MR compositing operations such as real-time passthrough, light weight person segmentation, and short-time recording are achievable on current Quest 3 hardware with careful optimization and adaptive quality control. However, more advanced features like high-quality real-time streaming or longer recording sessions may benefit from the computational headroom provided by next-generation SoCs like the Snapdragon 8 Gen 3 or a future XR-optimized variant. 

The approach creates a framework for iterative development and optimization that enables progressive improvement for MR capabilities with improving hardware and software optimization techniques. This approach maintains compatibility with current hardware and prepares the system for future improvements and expanded functionalities.

\begin{figure}[!ht]
\centering
\begin{tikzpicture}
    \begin{axis}[
        ybar,
        bar width=6pt,
        enlargelimits=0.05,
        ylabel={\%},
        ymin=0, ymax=70,
        xtick=data,
        symbolic x coords={
            Passthrough Rendering,
            Avatar + Scene Rendering,
            Depth Masking / Segmentation,
            720p Compositing + Encoding,
            Headroom
        },
        xticklabel style={rotate=30, anchor=east, font=\footnotesize},
        legend style={at={(0.5,-0.8)},anchor=north,legend columns=1, font=\scriptsize},
        nodes near coords,
        nodes near coords align={vertical},
        every node near coord/.append style={font=\scriptsize},
        width=8cm,   
        height=5cm   
    ]
    \addplot+[style={blue,fill=blue!30}] 
        coordinates {
            (Passthrough Rendering,12.5)
            (Avatar + Scene Rendering,32.5)
            (Depth Masking / Segmentation,22.5)
            (720p Compositing + Encoding,27.5)
            (Headroom,5)
        };
    \addplot+[style={orange,fill=orange!60}] 
        coordinates {
            (Passthrough Rendering,8.5)
            (Avatar + Scene Rendering,22.5)
            (Depth Masking / Segmentation,17.5)
            (720p Compositing + Encoding,17.5)
            (Headroom,34)
        };
    \legend{Snapdragon XR2 Gen 2 (Meta Quest 3), Snapdragon 8 Gen 3}
    \end{axis}
\end{tikzpicture}
\caption{Comparison of CPU and GPU Utilized in Different Pipeline Stages.}
\label{fig:resource-consumption}
\end{figure}
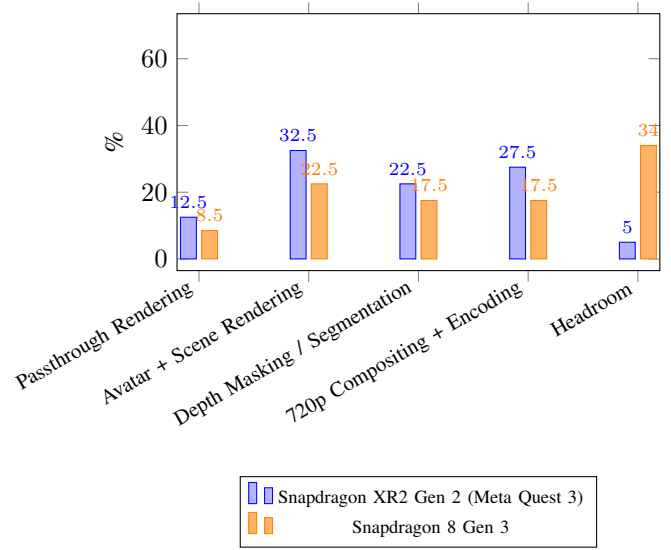

\section{Limitations}\label{sec:limitation}

This study presents several limitations that should be considered when interpreting the results and planning subsequent research. 

\subsection{Simulation-Driven Approach} The main bottleneck is that it depends on simulation-based performance estimation, rather than from real HW results. Simulation studies, although useful for early-stage system feasibility, can’t encompass real-world system complexities or precise cause-and-effect relationships in hardware-software interaction. While our estimates are based on existing standards and hardware specifications, they may not account for all environmental factors that can impact real system performance \cite{ieee2019hybridgpu,informs1997verification}. 

\subsection{Scopes of Validation} The feasibility is based on the hardware specs and benchmark data available in the public domain about the related ARM-based SoCs, such as Snapdragon 8 Gen 3 and MediaTek Dimensity 9300. In summary, due to proprietary firmware of Meta Quest 3 and unavailability of direct access to its native camera APIs, we cannot experimentally validate the proposed architecture, which may lead to possible inaccuracies of performance estimations \cite{nanoreview2024snapvsdim,semiengineering2024benchmarks}. 
 
\subsection{Limitations in Thermal Modeling} Projected thermal performance estimates are derived from generic mobile computing design principles, not from detailed modelling of the Quest 3 passive cooling system. In the real world the ambient temperature and user behavior provide a variability which cannot be fully addressed by the simulation framework \cite{ieee2018optic,informatics2025thermalpcm}. 
 
\subsection{Software Framework Dependencies} The proposed system relies on continued access and refinement of crucial software frameworks such as Unity Sentis for on-device ML inference and Meta’s experimental Passthrough Camera API. Changes in API availability or performance will affect the feasibility of the system \cite{unity2024sentis,meta2025pca}.

\subsection{Content Complexity Metrics} Performance is estimated for light and moderate scene complexity consistent with many of today's VR applications. Complex virtual worlds and avatars with many polygons will significantly increase the computational demands as well, limiting the available headroom for MR compositing \cite{milvus2025vrchallenges,arxiv2024cybersickness}. 
 
Together, these constraints emphasize the need for prototyping with Meta’s Passthrough Camera API in order to empirically verify simulated results, continued research in advanced thermal management to achieve sustained performance, and the consideration of next-generation XR SoCs for improved capabilities.

\section{Conclusion}\label{sec:summary}

In this work, we explore the feasibility of LIV SDK style MR compositing pipeline that can be implemented directly on standalone VR devices, we picked the Meta Quest 3 with the Snapdragon XR2 Gen 2 as for our target SoC. The work here was able to demonstrate that in Quest 3 the device possesses a capacity of performing short-term real time MR compositing 720p/30fps, both via simulation-based performance estimation and by relative hardware comparison. The optimization of the computational cost is crucial in order to be in line with thermal demands which are imposed by the passive cooling on the one hand.

The findings in this paper show that the XR2 Gen 2 chipset provides sufficient GPU and AI acceleration for lightweight passthrough rendering, avatar compositing and on-device ML segmentation, but it is found to be challenging that headset struggles with high-resolution and sustaining high frame rate MR capture potentially due to thermal constraints, and these aspects are expected to deserve more attention in future hardware. Our results underscore the importance of adaptive quality control and resource prioritization to enable acceptable performance given these limitations.

Compared to more powerful ARM-based SoCs, such as the Snapdragon 8 Gen 3 and MediaTek Dimensity 9300, we have room for improved MR compositing capability, towards future devices. However, these chipsets are not direct XR SoC replacements, generally these chipsets occupy more aggressive power envelopes that are not obvious fit for existing stand-alone VR thermal systems solutions and highlight the importance of hardware-software co-design in next-gen XR platforms. 

This research introduces an architecture and a simulation design to further support the development of native MR capture on standalone VR devices. The limitations that are revealed, namely reliance on simulators and no access to proprietary API, emphasize prototype implementation and empirical evaluation as a necessary immediate future work.

Additionally, future work should focus on porting ML-based segmentation models to XR hardware, advancing thermal management for prolonged recording and evaluating new SoC architectures like the Snapdragon XR2+ Gen 2 with limited additional improvement over the XR2 Gen 2. Moreover, the support to multi-platform and hybrid edge-cloud processing structuring are interesting development directions that can circumvent the limitations of current devices with respect to delivering richer and more spontaneous mixed reality contents.

In general, supporting native MR compositing on the latest generation of standalone VR hardware at low-resolution and frame rates is feasible but achieving high-quality continuous MR capture will require ongoing evolution of hardware, software optimization, and system-level innovation. By doing so, this paper provides a grounded view of the current possibilities and as well as a future roadmap for upcoming XR devices can come closer to achieve a fully integrated PC‐free mixed reality content creation experience.

\section{Future Work}\label{sec:future_work}

Expanding on these limitations, numerous future research directions are suggested. 

\subsubsection{Prototype Implementation and Validation} Implementing a working prototype using the Meta’s Passthrough Camera API is critical aspect to experimentally validate the simulation results and to refine performance models \cite{oculus2025samples}. 

\subsubsection{Advanced Thermal Management} Investigating dynamic thermal management techniques such as adaptive frame rate and resolution scaling can enable sustained performance for longer MR sessions \cite{acm2019heatpipe,tusacceptedpaper,xroffloadingtimescales}. 

\subsubsection{XR-Optimized and Lightweight ML Segmentation Models} It is important to design light, XR-optimized segmentation models for standalone VR devices. Techniques such as neural architecture search and model quantization also independently warrant further research \cite{unitysentisquestai,unity2024sentisdoc}. 

\subsubsection{Next Generation Hardware Evaluation}  With access to newer generation XR SoCs such as the Snapdragon XR3, The work investigating higher resolutions (i.e., 4K or 1080p60) and more advanced ML capabilities can continue to be pursued \cite{mlsys2023xrbenchrepeat}.

\subsubsection{Cross-Platform Support} Expanding the architecture to support multiple VR platforms and hardware configurations increases applicability and potential adoption across the entire XR spectrum.
 \cite{oculus2025samples}. 

\subsubsection{Hybrid Edge-Cloud System Architecture}  One interesting direction is to investigate hybrid processing models that can delegate computationally intensive MR tasks to the edge or cloud hosts over low-latency networks (e.g., 5G), driving MR experiences beyond device constraints \cite{xroffloadingtimescales,khan2024hybridoffloading}.


\end{document}